\documentclass[aps,pre,twocolumn,groupedaddress,showpacs]{revtex4}
\usepackage{graphicx}%
\usepackage{dcolumn}
\usepackage{amsmath}   
\usepackage{bm}
\usepackage{longtable}
\begin{document}

\title{\centering\Large\bf Model Energy Landscapes of Low-Temperature Fluids: Dipolar
                                  Hard Spheres } 
\author{Dmitry V.\ Matyushov}
\email[E-mail:]{dmitrym@asu.edu.}
\affiliation{Department of Physics and Astronomy and Center for Biophysics, 
  Arizona State University, PO Box 871604, Tempe, AZ 85287-1604}

\date{\today}
\begin{abstract}
  An analytical model of non-Gaussian energy landscape of low-temperature
  fluids is developed based on the thermodynamics of the fluid of
  dipolar hard spheres. The entire excitation profile of the liquid, from the
  high temperatures to the point of ideal-glass transition, has
  been obtained from the Monte Carlo simulations. The fluid of dipolar hard
  spheres loses stability when reaching the point of ideal-glass
  transition transforming via a first-order transition into a columnar
  liquid phase of dipolar chains locally arranged in a body-centered
  tetragonal order.
\end{abstract}
\maketitle

Thermodynamic and dynamic properties of supercooled liquids are often
related to the excess of configurations they possess relative to the
crystalline phase \cite{AngellJAP:00}. The number of states of the
liquid state is reflected by the configurational entropy $S_c(T)$
conventionally defined \cite{Lifshitz:78,Bryngelson:87,Freed:03} as the logarithm
of the density of states $\Omega(E)$ taken at the energy equal to the
internal energy $E(T)$ at temperature $T$:
\begin{equation}
  \label{eq:1}
   S_c(T) = \ln\left[\Omega(E(T))\right]. 
\end{equation}
The energy integral of the density of states and the Boltzmann factor 
provide the canonical configuration integral
\begin{equation}
  \label{eq:2}
  Z(\beta) = \int \Omega(E) e^{-\beta E} dE,
\end{equation}
where $\beta$ is the inverse temperature.

Stillinger \cite{Stillinger:82,Stillinger:88} gave an alternative
definition of the configurational entropy in terms of the density of
states $\Omega_{\phi}(E)$ of inherent structures, i.e.\ minima of the total
energy of the system as a function of its all translational an
rotational degrees of freedom (energy landscape). The canonical
configuration integral is now given by integrating over the basin
depths $\phi$ and the free energy of interbasin vibrations $F_v(\phi)$,
which is a weak, approximately linear, function of $\phi$
\cite{Buchner:99,Sastry:01}
\begin{equation}
  \label{eq:3}
   Z(\beta) = \int d\phi \Omega_{\phi}(\phi)e^{-\beta\phi -\beta F_v(\phi)} .
\end{equation}

Because of the central role of the configurational entropy in early
theories of viscous liquids by Adam, Gibbs, and DiMarzio
\cite{Gibbs:58,Adam:65}, and more recent random first-order transition
models by Wolynes and co-workers \cite{Xia:00}, much effort has been
invested in calculating the Stillinger configurational entropy,
$S_c^{\phi}(T)=\ln\left[\Omega_{\phi}(\phi(T))\right]$, from computer simulations of
model fluids \cite{Buchner:99,Sastry:01,Moreno:06}. A general
connection between $S_c(T)$ and $S_c^{\phi}(T)$ is, however, still unclear.

Analytical modeling of the landscape of supercooled liquids is to a
large extent based on the ideas advanced for spin glasses with
quenched disorder \cite{Mezard:87,Kirkpatrick:89}. The most prominent
role in application to structural glasses is played by Derrida's
random-energy, or Gaussian landscape, model (REM) \cite{Derrida:81}.
Computer simulations, mostly limited to temperatures above the
mode-coupling critical temperature, basically support the REM, in
particular the prediction of the $1/T$ falloff of the average basin
energy $\phi(T)$ from its high-temperature plateau
\cite{Buchner:99,Sastry:01}. However, general theoretical arguments
\cite{Stillinger:88,DMjcp5:05} and recent simulations \cite{Moreno:06}
suggest that the low-energy portion of the Gaussian landscape might be
inaccurate. In particular, combinatorial arguments suggest that the
derivative of the enumeration function, $\sigma_{\phi}(\phi) =
N^{-1}\ln[\Omega_{\phi}(\phi)]$, approaches infinity at the point of
ideal glass transition when the system runs out of configurations and
$\sigma_{\phi}(\phi_{IG})=0$ ($N$ is the number of liquid particles).
This infinite derivative eliminates the ideal glass transition at a
positive temperature \cite{Stillinger:88}. A phenomenological
description, patching together logarithmic and Gaussian enumeration
functions, was suggested to provide a correct low-energy portion of
$\sigma_{\phi}(\phi)$ \cite{Debenedetti:03}.

Generally, unlike the case of spin glasses, there has been a lack of
simple, solvable models of structural glasses. In this Letter we
propose a new model of non-Gaussian landscape of low-temperature
fluids with no quenched disorder. Our derivation is based on the
established thermodynamics of the model fluid of dipolar hard spheres
(DHS) \cite{Gubbins:84}. This monoatomic glass former (in contrast to
binary mixtures used in many recent simulations
\cite{Buchner:99,Sastry:01}) is long known to resist phase
transformations. This fluid lacks the liquid-vapor transition
decomposing instead into low-density dipolar chains
\cite{WeiPRL:93,Leeuwen:93}.  Upon cooling, dipolar hard spheres
transform into a ferroelectric fluid \cite{Wei:92}, but the
ferroelectric phase is stabilized by tin-coil boundary conditions used
in simulations and can be prevented by using a low dielectric constant
for the reaction-field or Ewald corrections for the cutoff of dipolar
interactions \cite{Wei:93}. This strategy has been employed in this
study. We have carried out the Monte Carlo (MC) simulations of the DHS
fluid within the standard NVT Metropolis protocol and the
reaction-field cutoff corrections. The reaction-field dielectric
constant of $\epsilon_{\text{RF}}=10$, below the lowest value $\simeq 18$
permitting the ferroelectric phase \cite{Wei:93}, has allowed us to
eliminate the transition to liquid ferroelectric.  The initial
configuration was set up as a face-centered cubic lattice of 108
dipolar spheres. The cubic simulation box also helps to suppress
crystallization of dipoles which do not favor highly symmetric lattice
structures \cite{Gao:00,Groh:01}.  The data were collected for
$(2-10)\times 10^7$ MC cycles.

Apart from avoiding crystallization, a significant advantage of the
DHS fluid is the existence of a simple analytical form for the free
energy $\beta F(\beta)= - \ln[Z(\beta)]$. The Pad\'e truncation of
perturbation series suggested by Stell and co-workers \cite{Larsen:77}
turned out to be very successful when tested against simulations
\cite{Gubbins:84}. The free energy of dipolar hard spheres
depends on two parameters, the reduced density $\rho^*=\rho\sigma^3$
and the reduced temperature $T=k_{\text{B}}T\sigma^3/m^2$.  Here,
$\rho$ is the number density, $\sigma$ is the hard-sphere diameter,
and $m$ is the dipole moment. All calculations and simulations here
have been done at constant volume with $\rho^*=0.8$ thus reducing the
number of variables to one.

Stell's Pad\'e solution for $F(\beta)$ is
\begin{equation}
  \label{eq:4}
  \beta F(\beta) = Nf(\eta) - \frac{A\beta^2}{1+b\beta } .
\end{equation}
Here $f(\eta)$ is the reduced free energy of the fluid of hard spheres
with zero dipole moment as a function of the packing density
$\eta=(\pi/6)\rho^*$.  For the Carnahan-Starling equation of state,
on has $f(\eta)=(4\eta-3\eta^2)/(1-\eta)^2$.  $\beta=1/T$ in Eq.\
(\ref{eq:4}) is given in reduced units and thus the coefficients
$A=Na$ and $b$ depend only on the liquid density through two-particle
and three-particle perturbation integrals,
$a=(\rho^*/6)I^{(2)}(\rho^*)$,
$b=(\rho^*/9)I^{(3)}(\rho^*)/I^{(2)}(\rho^*)$, tabulated by Larsen
\textit{et al} \cite{Larsen:77}.

\begin{figure}
  \centering
  \includegraphics*[width=5.5cm]{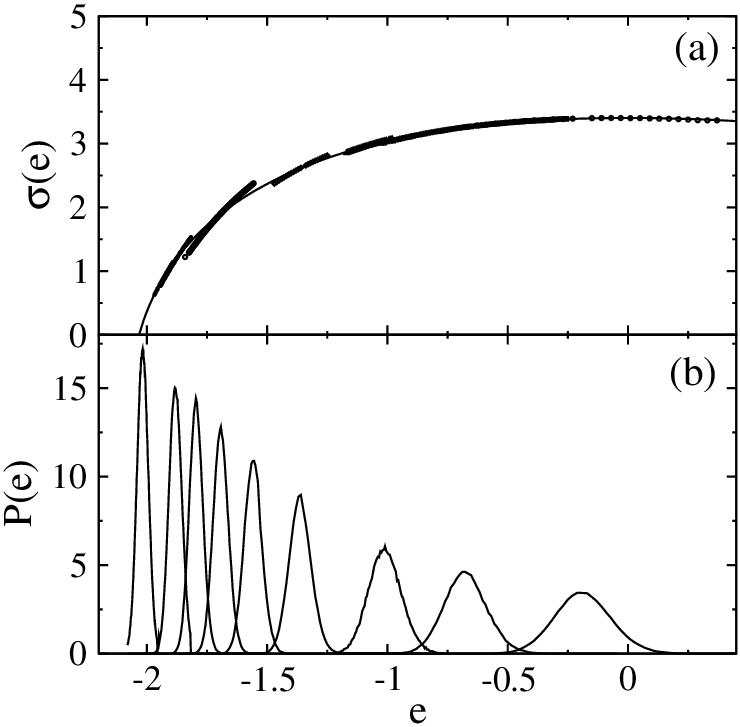}
  \caption{Enumeration function $\sigma(e)$ (a) and distribution function
    $P(e)$ (b) of the fluid of dipolar hard spheres at $\rho^*=0.8$ and
    the temperatures $T$ (from left to right in (b)): 0.125, 0.167,
    0.2, 0.25, 0.33, 0.5, 1.0, 2.0, 10.0. The simulation points in (a)
    are obtained from $P(e)$ within 95\% of the maximum probability at
    each temperature. }
  \label{fig:1}
\end{figure}

Following Freed \cite{Freed:03}, the density of states can be obtained
by inverse Laplace transformation of Eq.\ (\ref{eq:2}) in which we use
$\beta F(\beta)$ in the Pad\'e form given by Eq.\ (\ref{eq:4}).  The inverse
Laplace transform is calculated by expanding $\exp(-\beta F(\beta))$ in powers
of the second summand in Eq.\ (\ref{eq:4}) and performing pole
integration. The result can be converted to a closed-form equation:
\begin{equation}
  \label{eq:5}
  \begin{split}
  \Omega(e) = \left(b\sqrt{1+e/c}\right)^{-1}&\exp\left[N(f(\eta) -e/b - 2c/b)\right]\\
       &I_1\left(2(c/b)N\sqrt{1+e/c}\right), 
  \end{split}
\end{equation}
where $c= a/b$, $I_1(x)$ is the first-order modified Bessel function,
and $e=E\sigma^3 /Nm^2$ is the reduced internal energy per particle.
Since the argument of the Bessel function in Eq.\ (\ref{eq:5}) is
proportional to the number of particles $N$, its thermodynamic-limit
expansion gives the following expression for the enumeration function
$\sigma(e)=N^{-1}\ln[\Omega(e)]$:
\begin{equation}
  \label{eq:6}
  \sigma(e) = f(\eta) - b^{-1}\left(\sqrt{c + e} - \sqrt{c} \right)^2 .
\end{equation}

\begin{figure}
  \centering
  \includegraphics*[width=5.5cm]{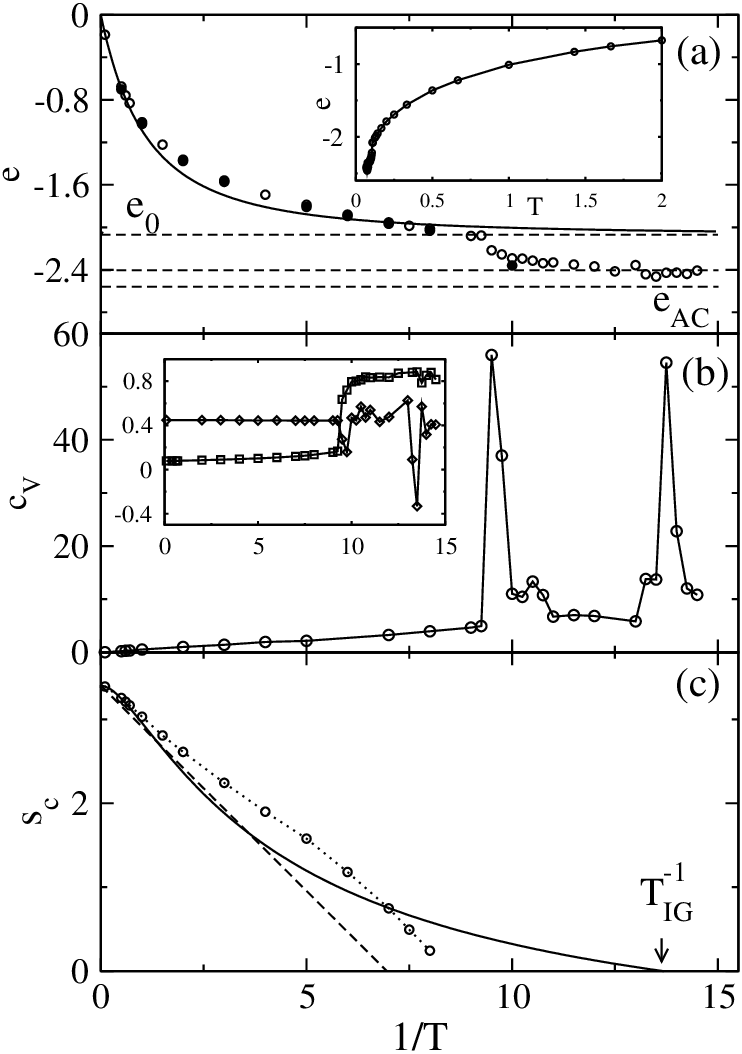}
  \caption{Average energy $e(T)$ (a), heat capacity $c_V$ (b), and the
    configurational entropy $s_c$ (c) vs $1/T$. The points in (a) are
    the results of MC simulations: open points for the internal
    energies and close points for the energies of inherent structures.
    The solid line in (a) shows the Pad\'e approximation of Eq.\
    (\ref{eq:7}). The dashed horizontal lines show the cutoff energy
    $e_0=-2.07$, the energy of non-interacting dipolar chains $e=-2.4$
    (middle line), and the energy of the closed-packed fcc
    antiferroelectric crystal $e_{AC}=-2.56$.  The inset in (a) shows
    $e(T)$ from simulations, the inset in (b) shows the nematic order
    parameter $P_2$ (squares) and the Binder parameter $M_B$
    (diamonds). The solid line in (c) is the configurational entropy
    $s_c(T)=\sigma(e(T))$ in the Pad\'e approximation [Eqs.\
    (\ref{eq:6}) and (\ref{eq:7})]. The dots show the configurational
    entropy [Eq.\ (\ref{eq:6})] with $e(T)$ from simulations. The
    dashed line in (c) is the thermodynamic excess entropy over that
    of the ideal gas calculated by temperature integration of
    simulated $c_V(T)$. }
  \label{fig:2}
\end{figure}

The enumeration function in Eq.\ (\ref{eq:6}) is Gaussian near its
top, $e\simeq 0$, where the dipole-dipole interactions are
insignificant and the limit of a hard-sphere fluid is reached. It
deviates from the parabolic shape in its low-energy wing, in
particular close to the low-energy cutoff at $e_0 = - c$ (Fig.\
\ref{fig:1}a). Depending on the parameters, Eq.\ (\ref{eq:6}) gives
two possible resolutions of the entropy crisis of low-temperature
fluids \cite{AngellJAP:00}.  When $f(\eta) - c/b \geq 0$, the
enumeration function has an infinite derivative at $e_0$ and,
according to Stllinger's arguments \cite{Stillinger:88}, there is no
ideal glass transition.  When $f(\eta) - c/b$ is strictly positive,
the entropy is non-zero at the cutoff energy, a situation observed for
network glass formers \cite{Saksaengwijit:04}. Finally, when $f(\eta)
- c/b < 0$, the ideal glass arrest $\sigma(e_{IG})=0$ happens before
the cutoff is reached, and the temperature of ideal-glass transition
$T_{IG}$, $e_{IG} = e(T_{IG})$ is positive. The use of
Carnahan-Starling hard-sphere free energy, and perturbation integrals
from Ref.\ \onlinecite{Larsen:77} gives $f(\eta) - c/b = -1.19$ at
$\rho^*=0.8$ and $T_{IG} \simeq 0.073$.

The enumeration function from Eq.\ (\ref{eq:6}) is compared to the
results of MC simulations in Fig.\ \ref{fig:1}a. The simulation points
were obtained by patching together central parts (within 95\% of the
maximum) of the distribution functions $P(e) \propto \exp(N(\sigma(e)
-\beta e))$ at different temperatures.  Since $P(e)$ defines
$\sigma(e)$ up to a constant, this comparison only serves to show the
non-Gaussian shape of the curve.  This non-Gaussian form of
$\sigma(e)$ forces the width of $P(e)$ to decrease with cooling (Fig.\
\ref{fig:1}b). This behavior is quite distinct from the prediction of
the Gaussian landscape model in which the width is
temperature-independent \cite{Derrida:81}.

The temperature dependence of the width is directly related to
the heat capacity since 
\begin{equation}
  \label{eq:9}
  P(e) \propto \exp\left[- \frac{(\delta e)^2}{2 c_V T^2} \right],
\end{equation}
where $\delta e$ is the energy fluctuation and $c_V$ is the excess
constant-volume heat capacity over that of the ideal gas.  The
high-temperature portion of the heat capacity, corresponding to the
distributions shown in Fig.\ \ref{fig:1}b, scales linearly with the
inverse temperature, $c_V \propto 1/T$ (Fig.\ \ref{fig:2}b). This
hyperbolic temperature scaling, often documented for structural glass
formers at constant pressure \cite{Richert:98}, results in a linear
temperature scaling of the squared width when the distribution $P(e)$
is fitted to a Gaussian function. This behavior was predicted by
models of supercooled liquids in terms of configurational excitations
\cite{DMjcp5:05}.

The calculation of the energies of inherent structures by conjugate
gradient minimization of configurations along simulated trajectories
results in energies $\phi(T)$ just slightly below $e(T)$ (closed points
in Fig.\ \ref{fig:2}a). This means that the long-range dipolar forces
produce essentially a mean-field potential for the local translations
and rotations and Eq.\ (\ref{eq:6}) for the density of internal
energies can be used for the density of inherent structures as well, $
\Omega(e) \simeq \Omega_{\phi}(e)$.

The density of states from simulations approaches the ideal-glass
state even steeper than Eq.\ (\ref{eq:6}) (Fig.\ \ref{fig:1}a). This
trend results in a slightly steeper temperature drop of $e(T)$ from
simulations compared to the result of the Pad\'e approximation (Fig.\
\ref{fig:2}a):
\begin{equation}
  \label{eq:7}
  e(T) = - c + c (1+b/T)^{-2} .
\end{equation}
In particular, instead of leveling off while approaching the cutoff
energy $e_0 = -2.07$, the system undergoes a first-order liquid-liquid
transition (see below) accompanied by a discontinuous drop of the
internal energy and a sharp peak in the heat capacity (Fig.\
\ref{fig:2}a,b). The orientational structure of the fluid also changes
as indicated by a step-wise increase in the nematic order parameter
$P_2$ calculated as the largest eigenvalue of the $\mathbf{Q}$-tensor
\begin{equation}
  \label{eq:8}
  \mathbf{Q} = (2N)^{-1}\sum_j(3\mathbf{\hat e}_j\mathbf{\hat e}_j -\mathbf{I}), 
\end{equation}
where $\mathbf{\hat e}_j$ is a unit vector along the dipole of
particle $j$.  The ferroelectric order parameter $P_1$ (normalized
total dipole moment of the liquid $\mathbf{M}$) remains close to zero
indicating that the liquid remains unpolarized. The Binder parameter
\cite{Challa:86}, $M_B = 1- \langle\mathbf{M}^4\rangle
/3\langle\mathbf{M}^2\rangle^2$, also shows downward spikes pointing
to first-order phase transitions (inset, Fig.\ \ref{fig:2}b).  The
internal energy at the first spike of $c_V$ drops from the level $e_0$
to the energy approximately equal to the energy of non-interacting
dipolar chains \cite{Tao:91} $e=-2.40$ (middle dashed line in Fig.\
\ref{fig:2}a).  This energy is slightly above the energy of
close-packed antiferroelectric fcc crystal $e_{AC}=-2.56$
\cite{Luttinger:46}.

The configurational entropy per particle $s_c(T)=\sigma(e(T))$ from
Pad\'e thermodynamics [Eqs.\ (\ref{eq:6}) and (\ref{eq:7})] deviates
from the hyperbolic functions at low temperatures approaching the
ideal glass transition at a positive temperature (Fig.\ \ref{fig:2}c).
In contrast, when $e(T)$ from simulations is substituted into Eq.\
(\ref{eq:6}), the decay of $s_c(T)$ is faster, reaching the
ideal-glass state near the point of the first-order transition. The
DHS fluids thus loses stability when it runs out of configurations
when approaching the ideal-glass transition and transforms into a
different thermodynamic state via a first-order transition.

\begin{figure}
  \centering
  \includegraphics*[width=5.5cm]{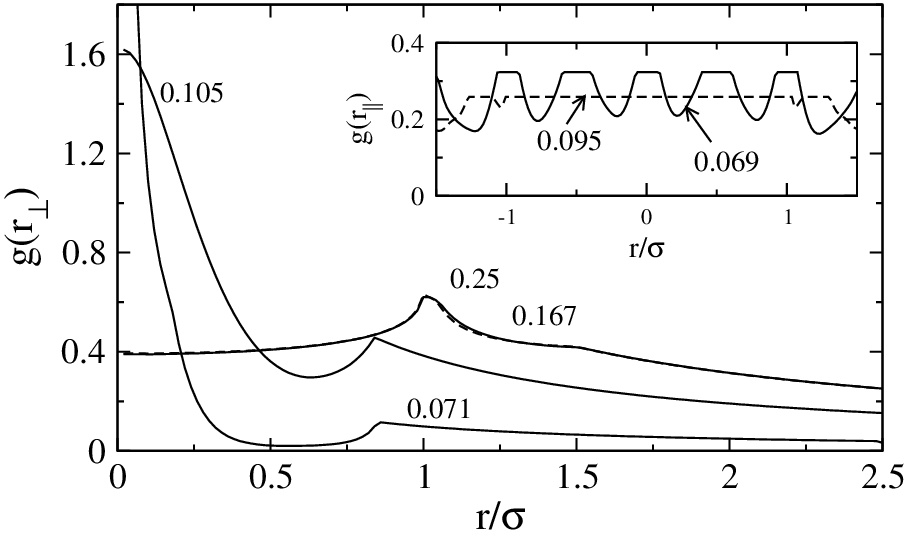}
  \caption{Pair distribution function of transverse interparticle
    separations (relative to the nematic director) at temperatures
    indicated in the plot. The high-temperature results for $T=0.167$
    (dashed line) and $T=0.25$ almost coincide on the scale of the
    plot. The distribution of distances parallel to the director in
    the inset shows the transition to the smectic phase of parallel
    chains with the local bct order.  }
  \label{fig:3}
\end{figure}

The properties of the low-temperature fluid phase are peculiar. Both
the density ($\propto \langle|\rho_k|^2\rangle$) and polarization
($\propto \langle |M_k|^2\rangle$) structure factors do not show
crystalline spatial or orientational order. However, snapshots of
simulated configurations (not shown here) indicate the existence of
chains of dipoles aligned head-to-tail.  These chains are persistent
for $(5-20)\times 10^3$ MC cycles resulting in the overall slow
convergence of simulations.  The chains of parallel dipoles are
arranged in bundles with the locally body-centered tetragonal (bct)
structure in which the two chains are displaced relative to each other
by the hard-sphere radius. This arrangement has the lowest energy for
fluids forming columnar phases of parallel dipolar chains
\cite{Tao:91,Hynninen:05}. A body-centered orthorombic lattice was
suggested as the ground state for Stockmayer ferroelectrics
\cite{Gao:00}.  In the present case, all the chain bundles are
oriented along the same director, but the net moment is compensated
among the oppositely oriented bundles.

The columnar structure is well seen in the pair correlation function
of transverse separations, which is very sensitive to columnar order
\cite{Wei:94} (Fig.\ \ref{fig:3}). At the phase transition, a peak at
zero transverse projection indicates the emergence of the columnar
phase. At the same time, a smaller peak, present at $r/ \sigma=1$ in
the isotropic dipolar fluid, shifts to $r/ \sigma = \sqrt{3}/2$
characteristic of the closest bct distance \cite{Tao:91}. The
one-dimensional dipolar chains experience strong Landau-Peierls
fluctuations with the orthogonal mean-square displacement scaling as
$\langle r_{\perp}^2\rangle \propto lT$ \cite{deGennes:70}, where $l$
is the average length of the chain. The fits of the initial portion of
$g(r_{\perp})$ result is almost invariant $\langle
r_{\perp}^2\rangle\simeq 0.07 \sigma^2$ suggesting that the length of
the chains increases as $1/T$.

The nematic columnar phase transforms into a smectic phase with
further cooling as indicated by the second peak of the heat capacity,
a downward spike of the Binder parameter (Fig.\ \ref{fig:2}b), and the
appearance of clear density modulation in the distribution function of
molecular separations parallel to the director (inset in Fig.\
\ref{fig:3}). 
 
In conclusion, a new model of thermodynamics of low-temperature fluids
offers a description of non-Gaussian landscape in a good agreement
with simulations. For the parameters of the DHS fluids studied here
the model predicts the excitation profile $e(T)$ to terminate at the
ideal-glass transition. The DHS fluid nearly avoids this point through
a first-order transition to a fluid phase with columnar order.

This research was supported by the the Air Force (FA9550-06-C-0084)
and NSF (CHE-0616646).

\bibliographystyle{apsrev}
\bibliography{/home/dmitry/p/bib/chem_abbr,/home/dmitry/p/bib/liquids,/home/dmitry/p/bib/glass,/home/dmitry/p/bib/dm,/home/dmitry/p/bib/dynamics,/home/dmitry/p/bib/ferro,/home/dmitry/p/bib/polymer}

\end{document}